\documentclass[amsmath,amssymb,aps,nofootinbib,showpacs,showkeys]{revtex4}
\usepackage[dvips]{graphicx}
\usepackage{graphicx}
\usepackage{amsfonts}
\usepackage{bm}
\usepackage{amsmath}
\usepackage{amssymb}
\usepackage{color}
\usepackage[all]{xy}

\newcommand{\rd}{\mathrm{d}} % Roman d for differential
\newcommand{\rF}{\mathrm{F}} 
\newcommand{\rG}{\mathrm{G}}

\begin{document}

\title{Chern Index of Taub-Bolt instantons in Einstein-Born-Infeld gravity}
\author{Daniel \surname{Flores-Alfonso}$^{1,}$}
\email[]{daniel.flores@correo.nucleares.unam.mx}
\author{Hernando \surname{Quevedo}$^{1,2,}$}
\email[]{quevedo@nucleares.unam.mx}
\affiliation{$^1$Instituto de Ciencias Nucleares,
Universidad Nacional Aut\'onoma de M\'exico,\\
AP 70543, Ciudad de M\'exico 04510, M\'exico \\
$^2$Department of Theoretical and Nuclear Physics,
Kazakh National University, 
Almaty 050040, Kazakhstan}

\begin{abstract}
Partition functions can be calculated by saddle point approximations whenever regular solutions exist, as they dominate the path integral. 
We examine a regular Taub-Bolt dyon of Einstein-Born-Infeld theory
which has electric and magnetic flux proportional to each other. Magnetic flux is found to be
essentially a winding number and the electric flux turns out to be indexed by it.%\\[0.2cm]
 %\textsl{MSC}: 32L81, 51P05, 55R15, 58J28, 81T70\\
 %\textsl{Keywords}: Einstein-Born-Infeld
\end{abstract}


\maketitle

\label{first}

\section{Motivation}

Gauge field theories can be  described by fiber bundles. In some cases,  the topological structure of these fiber bundles may have some physical significance.
The first such example  is 
Dirac's magnetic monopole \cite{dirac}  in which the interaction with electrons
leads to the quantization of the electric charge. Moreover, in the quantum Hall effect, the conductance of two-dimensional
electron systems has been observed to be quantized; in this case, the quantization is determined by the Chern number of the Berry curvature of the Brillouin manifold \cite{qhe}.

In gauge theories, winding numbers are commonly known to classify solutions. 
 This is especially true
for the magnetic flux, for instance, in the case of Bogomol'nyi-Prasad-Sommerfield (BPS) states, which are also analog to some gravity models with torsion \cite{canfora}.
Due to the geometric nature of torsion, there some  connections between gravity, high energy physics and solid state physics also appear as well.

In this work, we study the properties of dyons in curved spacetimes, corresponding to Abelian gauge fields with a U(1)$-$symmetry
that are defined over compact spaces with boundary. In a previous work \cite{fq1}, we explored Maxwell fields interacting with gravitational
instantons with a NUT parameter. In the case of Taub-Bolt spaces, the magnetic flux was found to be quantized
due to the topological structure of the underlying U(1)$-$bundle. Electric and magnetic fluxes are not independent
in these Einstein-Maxwell instanton as we must demand that the gauge field be regular on the entire manifold.
This  leads ultimately to an indexation of the electric flux. Recently, this result was  generalized to include the case of higher dimensions \cite{fq2}
and was illustrated by using Lovelock-Maxwell instantons. Here,  we investigate the four-dimensional Einstein
theory minimally coupled to Born-Infeld electrodynamics. We will prove that in this case the solutions are generalizations of the ones studied in \cite{fq1}.

\section{Born-Infeld Theory and Analytically Continued Solutions}

The sourceless equations of electrodynamics are
\begin{subequations}
 \begin{align}
  \rd F&=0,\label{f}\\
  \rd K&=0,\label{k}
 \end{align}
\end{subequations}
where the constitutive relation $K=K(F)$ in Born-Infeld theory is given by \cite{bi}
\begin{equation}
 K=\frac{\star F+\frac{\rG}{b^2} F}{\sqrt{1+\frac{\rF}{b^2}-\frac{\rG^2}{b^4}}}. \label{koff}
\end{equation}
Here $\rF=\langle F,F\rangle$ and $\rG=\langle F,\star F\rangle/2$. Notice that $K$ is nonlinear and, therefore, 
the theory is an example
of nonlinear electrodynamics, which is an exceptional theory as it does not present shock waves or birefringence.
In the limit $b\to\infty$, the constitutive relation becomes linear $K=\star F$.

We focus on the analytic continuation of the Born-Infeld source coupled to Einstein gravity, i.e., we will study the 
gravitational instantons of the theory. In particular, we will consider dyonic instantons with a NUT parameter, which 
were the first non-black-hole Euclidean solutions to be found.
The Lorentzian Einstein-Born-Infeld solution was found by Garc\'ia, Salazar and Pleba\'nski \cite{plebanski}
who generalize Brill's solution \cite{brill}. The geodesic motion on such a background has been recently studied 
by Bret\'on and Ram\'irez-Codiz in \cite{nora}.

We recall that gravitational instantons are used to approximate quantum gravity path integrals, which 
can be estimated through a saddle point approximation
\begin{equation}
 \int {\cal D}g e^{-I}=e^{-I[g_0]},
\end{equation}
where $I$ is the classical gravitational action and $g_0$ symbolizes a gravitational instanton.
A saddle point in the gravitational
scenario corresponds to a regular Euclidean solution, i.e., a solution that exists in the entire manifold. 
Consequently, a space with conical singularities cannot dominate the path integral. 
The Euclidean time circle which renders a black hole solution regular
at the horizon yields the Hawking temperature. In black holes, this temperature corresponds to the black body
radiation, which can be observed, in principle, outside the event horizon.
The temperature is also predicted by semiclassical gravity, where quantum fields propagate over 
a classical curved background.

\section{Taub-Bolt Instantons}

The general form of the Euclidean metric we consider is
\begin{equation}
 g=Y(r)\Delta\otimes\Delta+Y(r)^{-1}\rd r\otimes\rd r+(r^2-n^2)(\rd\theta\otimes\rd\theta+\sin^2\theta\rd\phi\otimes\rd\phi),\label{metric}
\end{equation}
where $n$ is the NUT parameter. For the precise form of $Y(r)$ we refer to \cite{nora}. 
For the metric to be defined globally, we must consider two regions with 
$\Delta_{\pm}=d\tau+2n(\pm1-\cos\theta) d\phi$.
Notice that the Euclidean time $\tau$ direction is Hopf fibered over a two-dimensional sphere. 
For any fixed value of $r$, the metric describes a quashed three$-$sphere, except at $r=r_+$ where $Y(r_+)=0$.
For the metric to be regular at $r=r_+$, the period of the Euclidean time $\tau$ must satisfy the condition 
$\Delta\tau=4\pi/Y'(r_+)=8\pi n$.
If $r_+$ coincides with the NUT parameter $n$, then the instanton is called Taub-NUT. In this case $r=n$ is called a nut
and is a regular point of the manifold. However, it is topologically trivial and so we focus on the complementary case.
When $r_+>n$, the instanton is called Taub-Bolt and the submanifold $r=r_+$ is called a bolt.

The field strength $F$ can be written locally as $F=dA$ with
\begin{equation}
 A=h(r)\Delta=\frac{1}{2n} [-q\sinh\Phi(r)+p\cosh\Phi(r)]\Delta , 
\end{equation}
where $\Phi(r)$ is an incomplete elliptic integral of the first kind such that $\Phi(\infty)=0$ (cf. \cite{nora}).
Since the Euclidean time direction is degenerate at the bolt, then we must have $h(r_+)=0$ for the gauge field to be regular everywhere
on the manifold. Except for the explicit form of $h(r)$, the field strength $F$ has the same form as in \cite{brill} and, therefore, we will say that it is of Brill type \cite{fq2}.

The Chern number of the above configuration is
\begin{equation}
 c_1\smile c_1[M]=cs[\partial M]=\frac{1}{4\pi^2}\int\limits_{S^3_{\infty}}A\wedge F=(2p)^2.
\end{equation}
The invariant only depends on the field at infinity. Notice that at the boundary
\begin{equation}
 A^{\infty}_{\pm}=h(\infty)\Delta_{\pm},
\end{equation}
so it follows that
\begin{equation}
 A_+-A_-=2p\rd\phi.
\end{equation}

Since U(1) is an Abelian group, we can read off the transition function which characterizes the principal bundle.
The transition function maps the intersection of the covers, where the gauge potentials are well defined, into the gauge group.
Notice that the intersection has the homotopy type of a circle parametrized by $\phi$.
The Brouwer degree of this map is $\omega=2p$, which is an integer representing the number of times the intersection circle
winds around U(1). This ultimately determines the topology of the bundle space. The Chern number is $c_1^2=\omega^2$
and the magnetic flux is
\begin{equation}
 \int\limits_{S^2_{\infty}} F=2\pi\omega,\qquad\omega\in\mathbb{Z}.\label{mflux}
\end{equation}
Furthermore, the electric flux is given by
\begin{equation}
 \int\limits_{S^2_{\infty}}K=-4\pi q=-2\pi\coth\Phi(r_+)\omega,\qquad\omega\in\mathbb{Z}.\label{eflux}
\end{equation}
Notice that we include the asymptotic boundary $\partial M$ in the manifold $M$ and so it is compact.
As pointed out in \cite{ao}, the second order homology group of the entire space $H_2(M,\mathbb{Z})$
classifies the magnetic flux. Since the Betti number $b_2(M)=1$, there is exactly one independent
component of the magnetic flux that is quantized. A similar argument applies for the electric flux so that it becomes classified by 
the homology of the boundary. In this case $b_2(\partial M)=0$ and so the homological approach tells us that there
is no quantization of the electric flux. However, formula (\ref{eflux}) is a result not of the topology directly, as (\ref{mflux}) for example,
but of the regularity condition imposed on the instanton. Let us recall  that this assumption 
is done in order for the saddle point approximation to hold.

Upon examination of the Ricci scalar \cite{plebanski,nora}, it is straightforward to see that
Taub-NUT instantons without curvature singularities are only possible if $p^2-q^2=0$.
This choice collapses the Born-Infeld solution into Maxwell's. Recall that in dimension four
Taub-NUT instantons have selfdual U(1) fields (see, e.g., \cite{fq1}). This also entails that there are no
Eguchi-Hanson type solutions in Einstein-Born-Infeld theory. This is due to
the essential selfdual property of the electromagnetic fields defined on the background.
The non-selfdual nature of the Taub-Bolt solutions makes them the basic example of Einstein-Born-Infeld instantons
with SU(2) isometry.

\section*{Acknowledgements}
DFA was supported by the CONACyT Grant with No. 404449. 
This work has been supported by the UNAM-DGAPA-PAPIIT, Grant No. IN111617.

\label{last}

\end{document}